\documentstyle[11pt,moriond,epsfig,bbox]{article}

\def\Journal#1#2#3#4{{#1} {\bf #2}, #3 (#4)}

\def\PRC{{\em Phys. Rev.} C}

\def\PREP{\em Phys. Rep.}

\def\be{\begin{equation}}
\def\ee{\end{equation}}
\def\bea{\begin{eqnarray}}
\def\eea{\end{eqnarray}}

\def\bq{\bbox{q}}
\def\bK{\bbox{K}}
\def\bp{\bbox{p}}

\begin{document}
\vspace*{3.7cm}
\title{BOSE-EINSTEIN CORRELATIONS IN HEAVY-ION PHYSICS \\
       AND ELECTRON-POSITRON COLLISIONS}
\author{Ulrich Heinz$^{a,b}$ and Pierre Scotto$^b$}

\address{$^a$Theoretical Physics Division, CERN, CH-1211 Geneva 23, 
         Switzerland\\
         Institut f\"ur Theoretische Physik, Universit\"at Regensburg,
         D-93040 Regensburg, Germany}

\maketitle\abstracts{
We shortly review recent successes in applying Bose-Einstein 
interferometry in heavy ion collisions and the proceed to some model
calculations for 3-dimensional Bose-Einstein correlation functions
in $e^+e^-$ collisions at the $Z^0$ pole.}

\section{Theoretical Overview}

Bose-Einstein correlations (BEC) are a phase-space phenomenon: 
Symmetrization of the multiparticle wave function affects the 
measured $n$-particle coincidence spectra and leads to an 
enhancement relative to the corresponding product of independent 
1-particle spectra, if the emitted particles are close in 
phase-space (i.e. they occupy the same elementary phase-space 
cell). The spatial length of the elementary phase-space cells 
is limited by the geometric size of the source of particles 
with the considered momentum. The larger this size, the narrower 
these cells are in momentum space. By tuning the relative momenta 
and watching the onset of BEC effects one can thus measure the 
spatial length of the elementary phase-space cells and thereby 
the size of the source.


{\it Wigner Functions.} A description of BEC effects among $n$
particles thus involves the $n$-particle phase-space density. Since we
are discussing a quantum mechanical phenomenon, we are not talking
about a classical phase-space density (which has directly a
probabilistic interpretation), but about the Wigner density (which is
positive definite only when averaged over many elementary phase-space
cells). If the particles are emitted independently, the
(unsymmetrized) $n$-particle Wigner density factorizes, and all
$n$-particle coincidence cross sections are expressible through the
single-particle Wigner function $S(x,p)$. The assumption of
independent particle emission is justifiable in heavy ion collisions
where the many unobserved particles serve as a reservoir for all kinds
of conserved quantities. In $e^+e^-$ collisions this is much less
obvious and needs to be tested experimentally. 


{\it Correlation Function.} As long as the source has sufficiently low
phase-space density that multi-particle symmetrization effects are
dominated by two-particle exchange terms, the two-particle correlation
function $C(\bq,\bK)$, defined as the ratio of the 2-particle
coincidence spectrum $P_2(\bp_a,\bp_b)$ and the product of
single-particle spectra $P_1(\bp_a) P_1(\bp_b)$ with
$\bq{\,=\,}\bp_a{-}\bp_b$ and $\bK{\,=\,}(\bp_a{+}\bp_b)/2$, is given
by\,\cite{WH99}\,\footnote{We here neglect Coulomb final state
  interactions since methods are known to correct the data for
  them.\cite{WH99}} 
 \be
 \label{1}
   C(\bq,\bK) = {\cal N} \left( 1 + 
   {\vert\int_x S(x,K)\,e^{iq{\cdot}x}\vert^2 \over
    \int_x S(x,p_a)\, \int_x \, S(y,p_b)}\right)
   = {\cal N} \left( 1 + {P_1(\bK)^2\over P_1(\bp_a) P_1(\bp_b)}
      \left\vert{\int_x S(x,K)\,e^{iq{\cdot}x} \over
                 \int_x S(x,K)}\right\vert^2\right)\,.
 \ee
Here $\int_x\equiv\int d^4x$, $q^0{\,=\,}E_a{-}E_b$, 
$K^0{\,=\,}(E_a{+}E_b)/2$, and 
 \be
 \label{2}
   P_1(\bp) = \int_x S(x,p) \quad {\rm with} \quad 
   p^0=E_p=\sqrt{m^2+\bp^2}\,.
 \ee
The normalization ${\cal N}$ depends on the multiplicity 
distribution via\,\cite{ZSH98} ${\cal N} = \langle n(n{-}1) 
\rangle/\langle n\rangle^2$. In heavy ion collisions usually 
${\cal N}\approx 1$. Due to the mass-shell constraint\,\cite{WH99} 
$q^0{\,=\,}\bbox{\beta}\cdot\bq$ (where 
$\beta{\,=\,}\bK/K^0{\,\approx\,}\bK/E_K$ is the velocity of the 
particle pair) the Fourier transform in (\ref{1}) is not invertible: 
the separation of temporal and spatial aspects of the emission 
function $S(x,K)$ requires additional model assumptions which must 
be provided by a physical picture of the time evolution of the source 
until freeze-out.\cite{WH99}


{\it The Reduced Correlator}. While (\ref{1}) goes to $2{\cal N}$ at
$\bq{\,=\,}0$, real correlation functions usually approach a smaller
value ${\cal N}(1{+}\lambda)$ with $\lambda(\bK) < 1$. Possible 
reasons are partial phase coherence in the source and decay
contributions from long-lived resonances.\cite{WH99} To account for
this one rewrites (\ref{1}) as
 \be
 \label{3}
   C(\bq,\bK) = {\cal N} \Bigl( 1 + \lambda(\bK) {\cal K}(\bq,\bK)\Bigr)
   ={\cal N} \left( 1 + \lambda(\bK) 
    {P_1(\bK)^2\over P_1(\bp_a) P_1(\bp_b)} 
    {\cal K}_{\rm red}(\bq,\bK)\right)\,.
 \ee
The {\em reduced correlator} ${\cal K}_{\rm red}(\bq,\bK)$ is given 
by the last term in (\ref{1}) which contains the information about 
the space-time structure of $S(x,K)$. To isolate it one constructs 
$C(\bq,\bK)$ from the measured 1- and 2-particle cross sections, 
applying the Coulomb correction, determines ${\cal N}$ and 
$\lambda(\bK)$ from the limits $q{\,\to\,}0$ and $q{\,\to\,}\infty$,
divides by ${\cal N}$ and subtracts the 1, and finally divides the 
result by $\lambda(\bK)$ and the measured ratio of single particle 
cross sections $P_1(\bK)^2/P_1(\bp_a) P_1(\bp_b)$. For large sources 
like those in heavy ion collisions this ratio is close to unity,\cite{CSH95} 
but for small sources like those in $e^+e^-$ it can contribute 
significantly to the $\bq$-dependence of $C(\bq,\bK)$; it is then 
important to divide it out before trying to extract the source size. 
{\em So far we have seen no data analysis where this is done!} 
Instead, one usually extracts the size directly from 
${\cal K}(\bq,\bK)$, without dividing out the 1-particle 
spectra. As we will see, this can be quite misleading.


{\it Source Radii from BEC.} One usually characterizes\,\cite{WH99}
the source function $S(x,K)$ by its norm, center and space-time
variances (widths), all of which are generally functions of the
momentum $\bK$ of the emitted particles. In this ``Gaussian
approximation'' the reduced correlator reads
 \be
 \label{5}
   {\cal K}_{\rm red}(\bq,\bK) = \exp\Bigl[ - q^\mu q^\nu 
   \langle \tilde x_\mu \tilde x_\nu\rangle(\bK)\Bigr] \,,
 \ee
where $\langle \tilde x_\mu \tilde x_\nu\rangle =
\langle x_\mu x_\nu\rangle - \langle x_\mu \rangle
\langle x_\nu\rangle$, with
 \be
 \label{6}
   \langle x_\mu x_\nu\rangle(\bK) = 
   {\int_x x_\mu x_\nu\, S(x,K)\over \int_x S(x,K)}\,,
 \ee
are the space-time variances of the emission function (effective source
sizes). Different conventions for resolving the mass-shell constraint
$q^0=\bbox{\beta}\cdot\bq$ and expressing (\ref{5}) in terms of three
indepenent components of $q$ lead to different Gaussian parametrizations
for the correlator.\cite{WH99} The corresponding Gaussian width 
parameters, the ``HBT (Hanbury Brown - Twiss) radii'', are then 
combinations of the variances $\langle \tilde x_\mu \tilde 
x_\nu\rangle(\bK)$ and thus functions of the pair momentum $\bK$.

\section{Bose-Einstein Correlations in Heavy Ion Collisions}

Due to space reasons we will be very short -- detailed discussions 
can be found elsewhere.\cite{WH99,TWH99} For Pb+Pb collisions at 
the SPS it was found that the pion emitting source is a rapidly 
expanding fireball in approximate local thermal equilibrium which
at decoupling has a temperature of about 100 MeV and expands nearly
boost-invariantly in the longitudinal direction while the average 
transverse expansion velocity is a bit larger than half the light 
velocity. The collective expansion manifests itself in a strong
and characteristic dependence of the space-time variances $\langle 
\tilde x_\mu \tilde x_\nu\rangle$ of the effective source $S(x,K)$
on the pair momentum $\bK$. This implies a corresponding 
$\bK$-dependence of the HBT radii extracted from (\ref{5}). 
The pion emission process lasts only for about 2-3 fm/$c$ but 
it doesn't begin until at least 6-8 fm/$c$ after the collision.
Freeze-out thus is a rather sudden process at the end of an 
extended rescattering and expansion stage. It is important to 
stress that the separation of longitudinal and transverse flow and 
access to the emission duration $\langle \tilde t^2\rangle$ is only
possible in a full-fledged 3-dimensional and $\bK$-dependent analysis 
of the correlation function $C(\bq,\bK)$. Projections to lower 
dimensionality (e.g on $q_{\rm inv}^2$) lead to uncontrollable 
and unrecoverable loss of information. 

\section{Bose-Einstein Correlations in $e^+e^-$ Collisions}

As stated in Sec. 1, to compute Bose-Einstein correlations one 
needs information on the Wigner phase-space density of the source.
Going simulation programs of particle production in high-energy
$e^+e^-$ collisions like PYTHIA, JETSET and HERWIG provide only 
momentum-space information on the produced particles. This is 
not enough to calculate BEC effects. Different methods have been 
suggested to provide the missing coordinate-space information, 
either directly or indirectly.\cite{BEMC} We previously 
studied\,\cite{GEHW00} BEC in VNI which studies the time evolution
of the collision in phase-space. Here we present some very early 
results based on a phase-space version\,\cite{Sharka} of JETSET~7.4
which provides both the momenta and production coordinates for
the produced particles. Our version of this code distributes the
transverse distance of the production points from the central string
axis according to a Gaussian with rms radius of 0.78\,fm while
S.~Todorovova's version\,\cite{Sharka} puts the production points
right on the string axis. This latter procedure is inconsistent with
the uncertainty relation, and we found accordingly\,\cite{ZWSH}
that it produces correlation functions which rise as a function of
$q_s$ instead of decaying.

The algorithm for computing the correlation function from the
positions and momenta of the generated pions is described 
elsewhere;\cite{GEHW00} we use the ``classical'' algorithm 
without wave packet smearing.\cite{GEHW00} In order to test the
space-time structure of the events generated by JETSET and the BEC
afterburner, we begin with a simple event topology ($e^+e^-\to Z^0 
\to q\bar q\to 2$\,jets) and consider only directly produced pions,
thus avoiding the multiscale problems associated with longlived
resonance decays. We analyse the correlation function in a Cartesian
coordinate system where the longitudinal ($l$ or L) axis is along the 
direction defined by the relative momentum of the initial $q\bar q$
pair ($\approx$ jet axis), the outward ($o$ or T) direction is defined
by the transverse pair momentum $\bK_{\rm T}$, and the sideward ($s$)
axis points in the third direction.
 
\begin{figure} [ht]
\epsfig{file=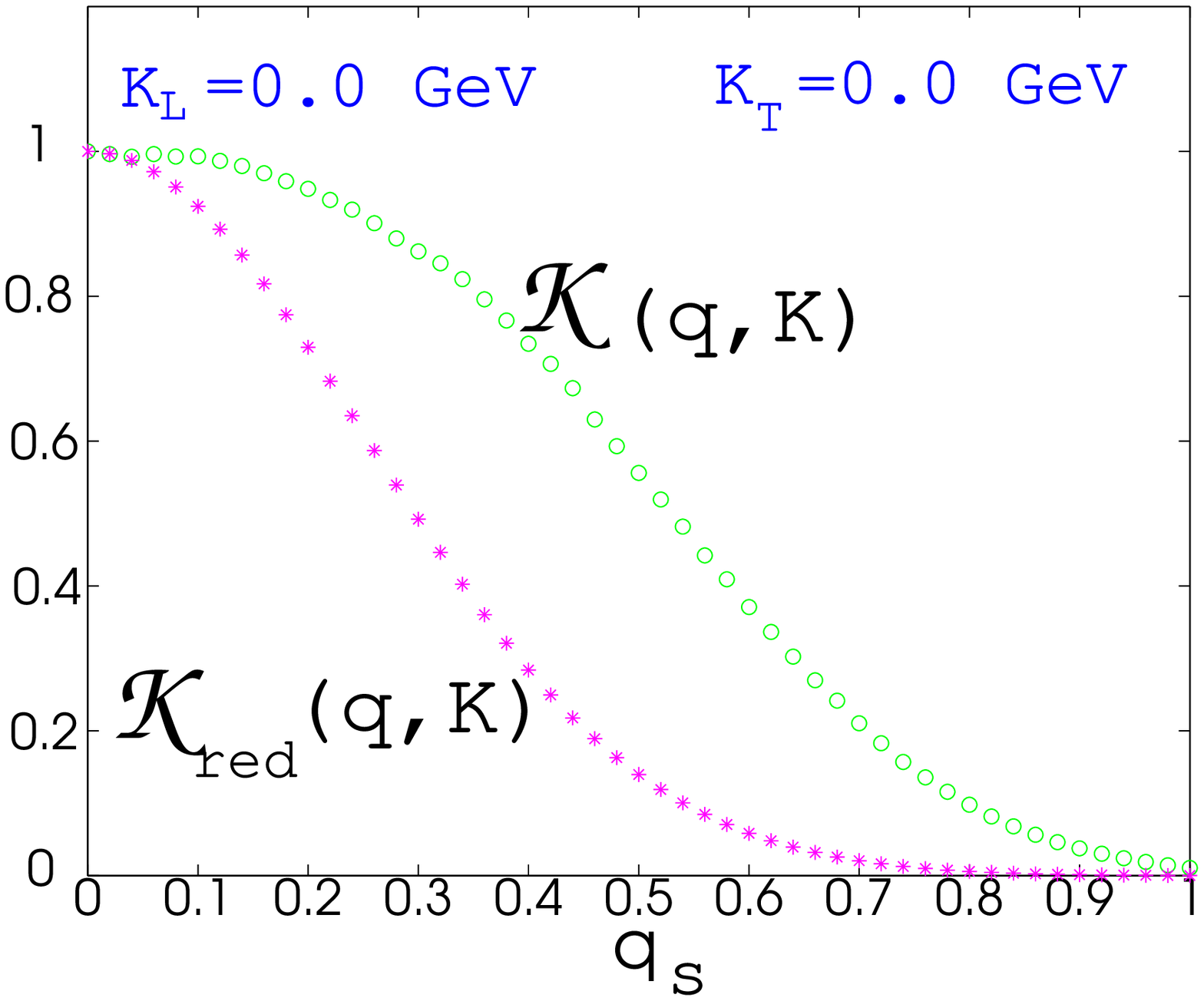,width=5.2cm}
\epsfig{file=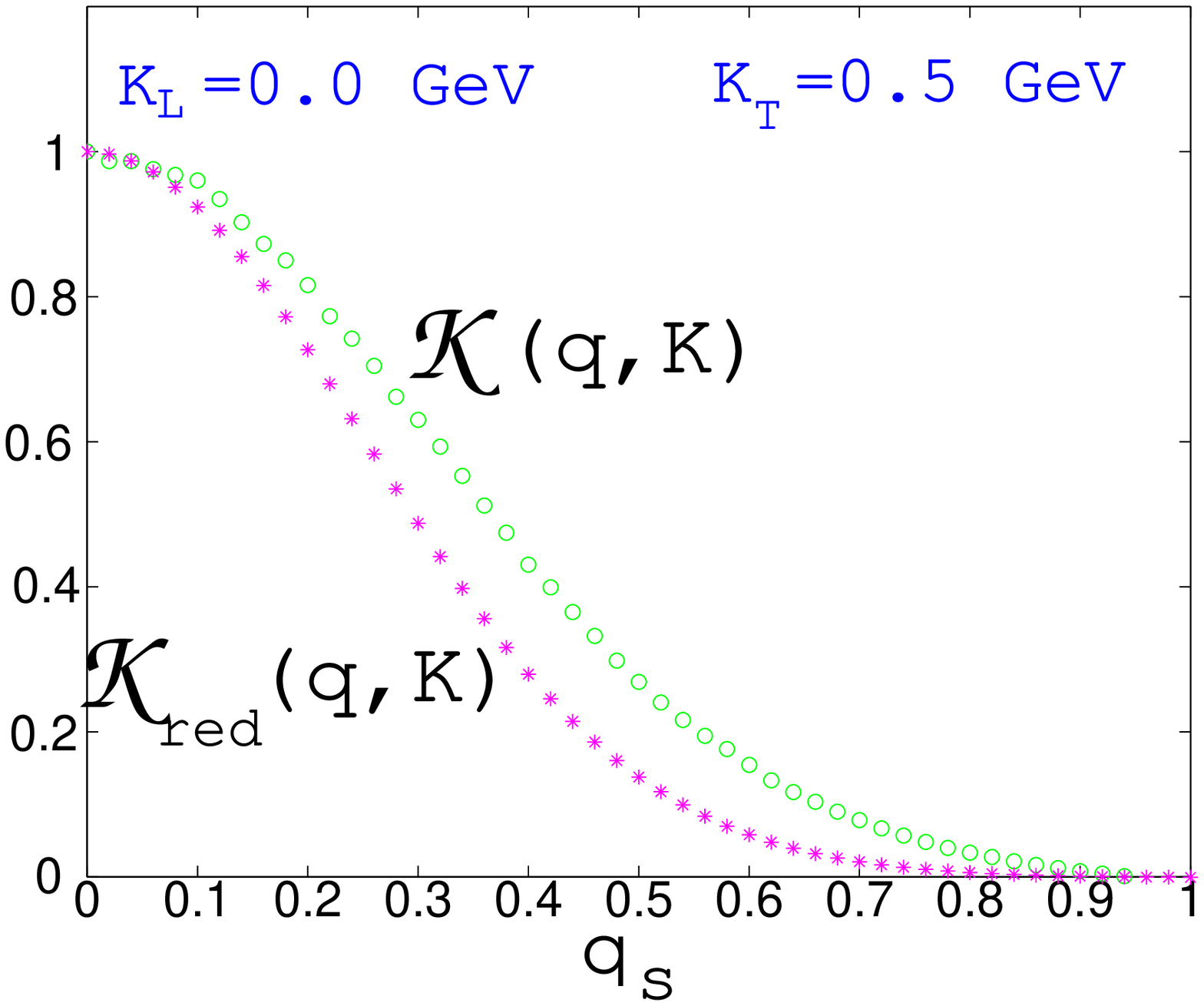,width=5.2cm}
\epsfig{file=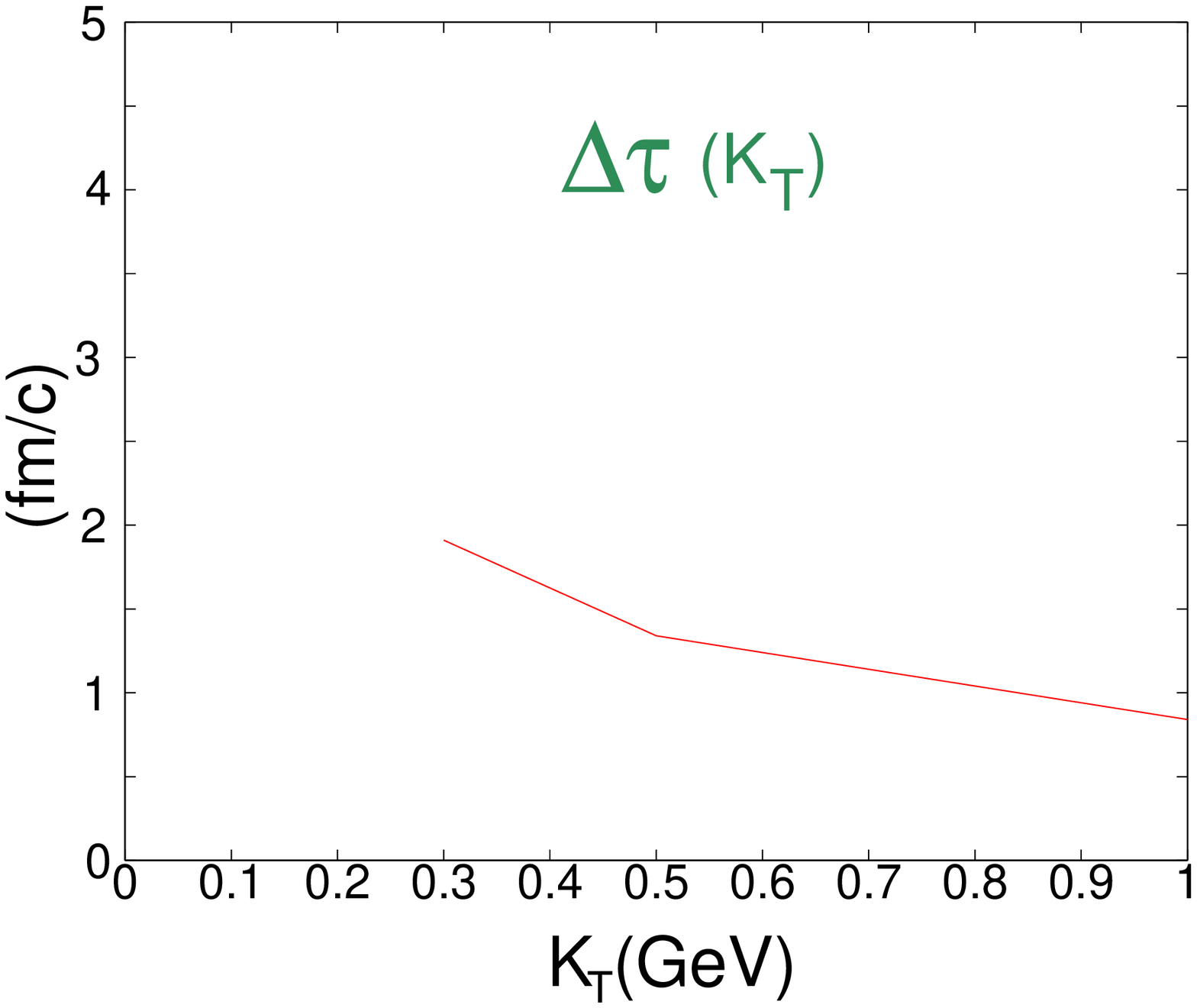,width=5.2cm}\\
\epsfig{file=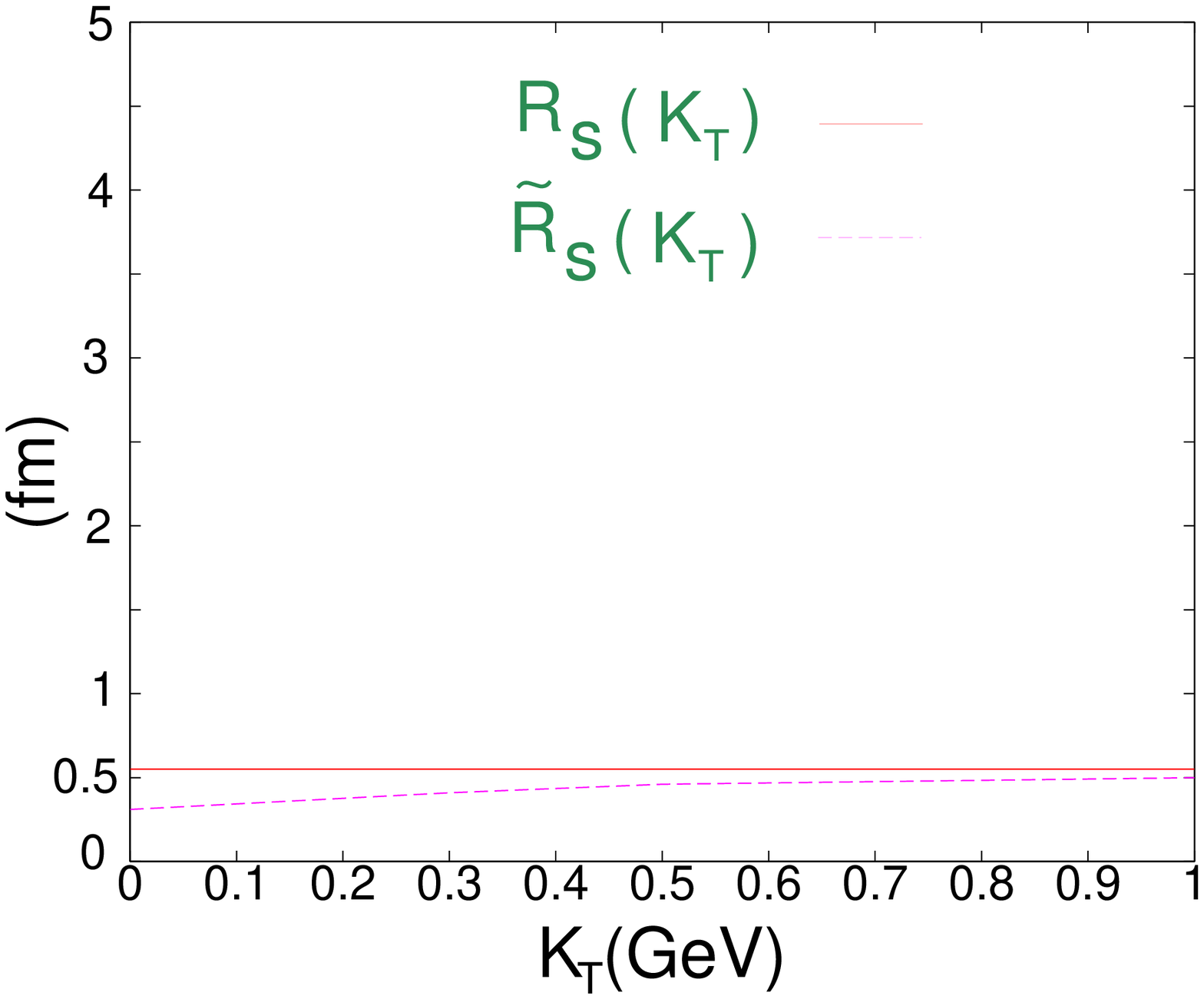,width=5.2cm}
\epsfig{file=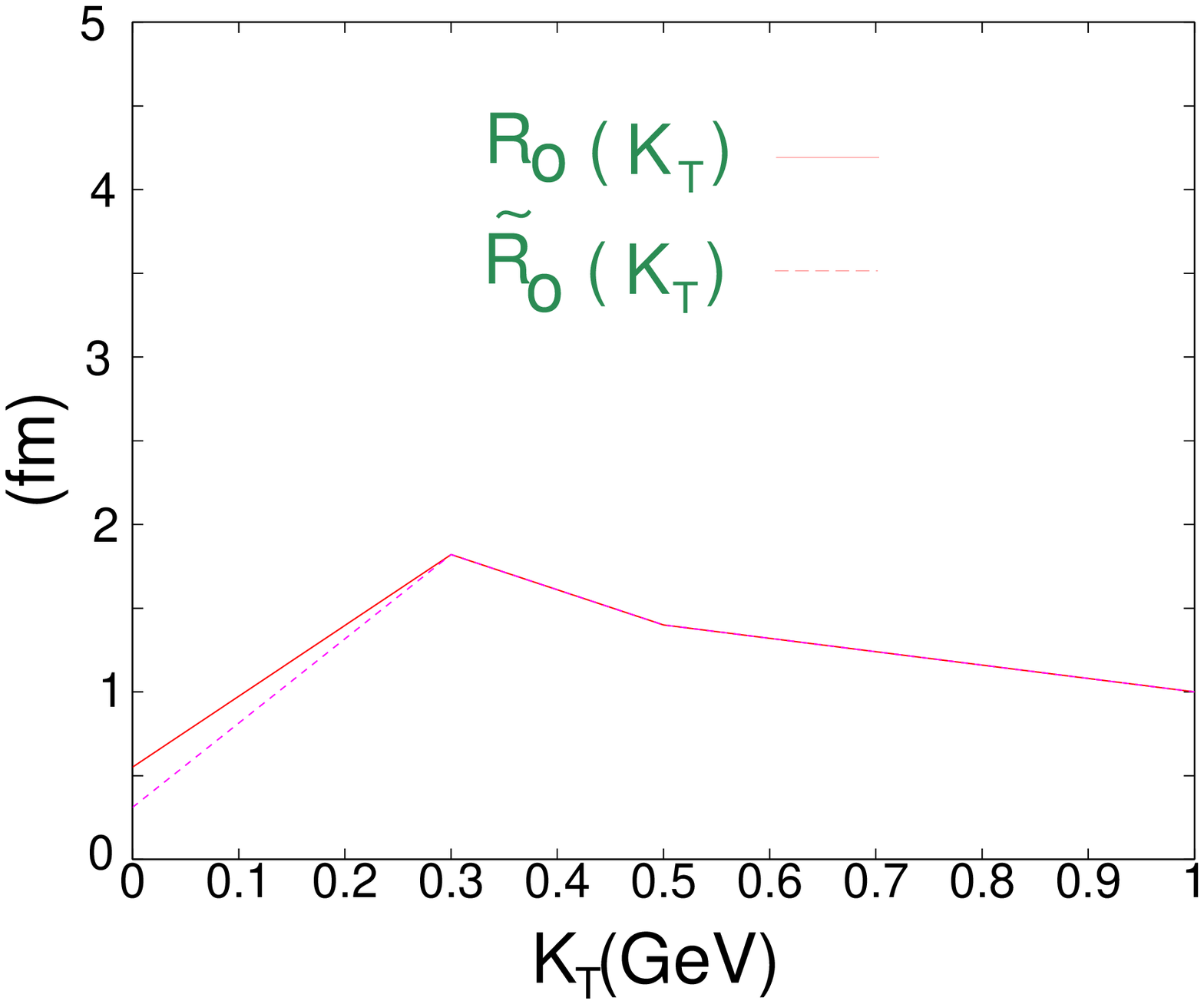,width=5.2cm}
\epsfig{file=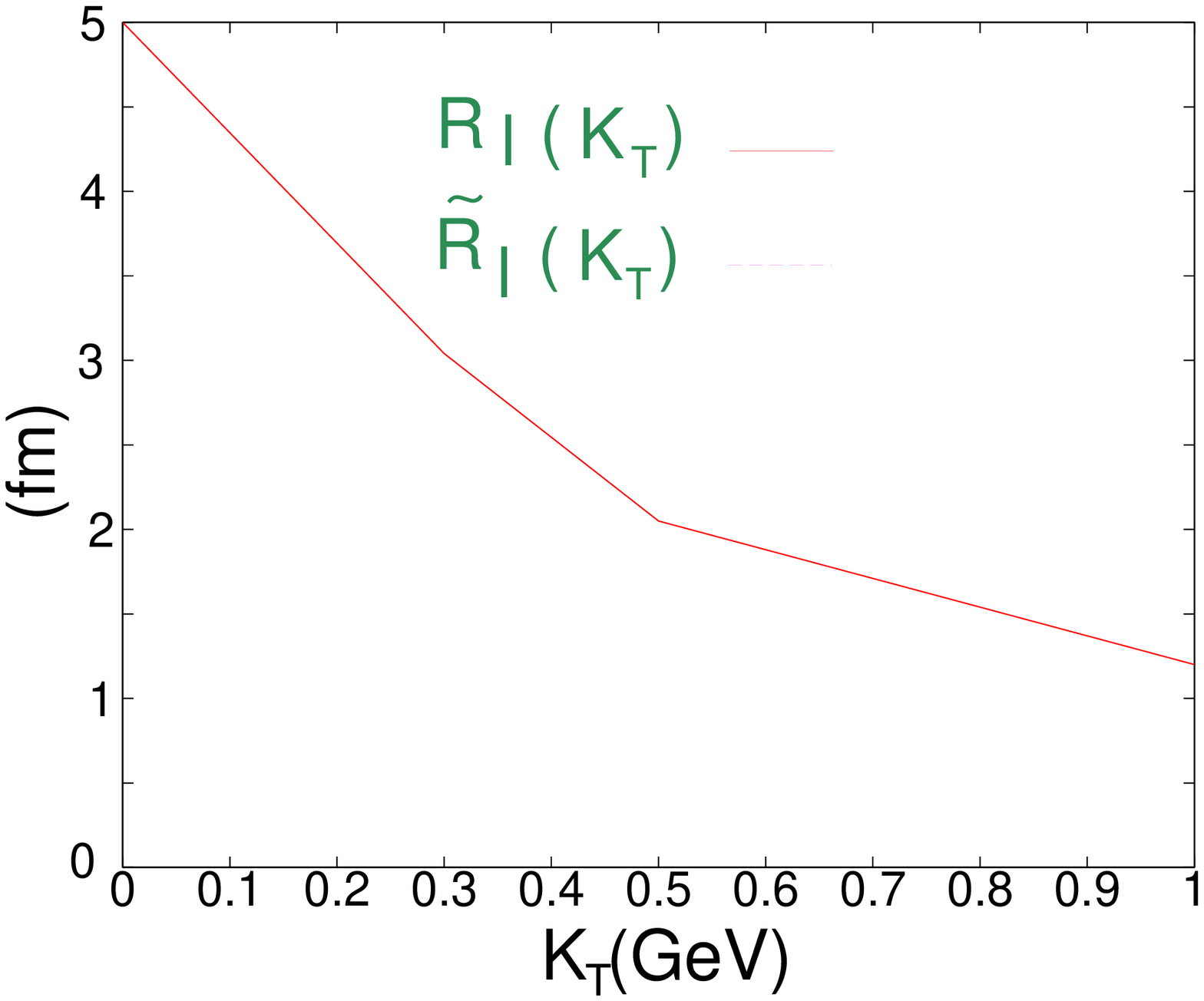,width=5.2cm}
\caption{First two panels: The correlation function in side-direction
    ($q_o{\,=\,}q_l{\,=\,}0$) for pairs with $K_{\rm L}{\,=\,}0$ (such 
    that the $R_{ol}$ cross term vanishes\,\protect\cite{WH99}) and 
    $K_{\rm T}{\,=\,}0$ and 0.5\,GeV, respectively. Third panel: The
    emission time duration 
    $\Delta\tau{\,=\,}\sqrt{R_o^2-R_s^2}/\beta_{\rm T}$ as a function
    of $K_{\rm T}$ for $K_{\rm L}{\,=\,}0$. Second row: $R_s$, $R_o$,
    and $R_l$ as functions of $K_{\rm T}$ for $K_{\rm L}{\,=\,}0$. We
    checked that the HBT radii correctly reproduce the rms widths
    of the space-time scatter plots of the produced pions in the
    appropriate $\bK$-windows.
\label{F1}}
\end{figure}

The top two left panels of Fig.\,\ref{F1} show the correlator in the
side direction. The reduced correlator ${\cal K}_{\rm red}$ is seen to
be independent of $K_{\rm T}$ and always reproduces the input rms
width of the string: $R_s{\,=\,}r_{\rm rms}/
\sqrt{2}{\,=\,}0.55$\,fm. In contrast, ${\cal K}$ does depend on   
$K_{\rm T}$, and for small $K_{\rm T}$ it produces smaller HBT radii
(0.31, 0.41, 0.46 and 0.50\,fm at $K_{\rm T}{\,=\,}0$, 0.3, 0.5 and
1.0\,GeV, respectively). This effect is an artifact induced by the 
ratio of 1-particle spectra in (\ref{3}); it matters since the real
radius is so small, producing significant errors if not divided
out. For $R_o$ and $R_l$, on the other hand, its effect is in our
calculation nearly negligible: these radii come out much larger than
$R_s$. This, however, points to another problem: longitudinal HBT
radii of up to 5\,fm are incompatible with the data which give only
about 1\,fm (see the experimental talks in this session)! The problem
seems to be connected with the large emission time duration
$\Delta\tau$ of up to 3 fm/$c$ at low $K_{\rm T}$. This parameter,
which reflects the proper time distribution of string breaking
processes in JETSET, is not fixed by 1-particle spectra, but it is
seen to seriously affect the 2-particle correlations. We are presently 
trying to fix this problem. At this moment we can only say that the
version of JETSET used by us disagrees with experiment at the level of 
2-particle correlations.

\noindent{\bf Ackowledgement}: This work was supported by the Deutsche
Forschungsgemeinschaft.


\end{document}